\documentclass[letterpaper,english,twocolumn,superscriptaddress]{revtex4-1}
\usepackage[T1]{fontenc}
\usepackage[latin9]{inputenc}
\usepackage{babel}
\usepackage{textcomp}
\usepackage{amsmath}
\usepackage{amssymb}
\usepackage{graphicx}
\usepackage[unicode=true,pdfusetitle,
 bookmarks=true,bookmarksnumbered=false,bookmarksopen=false,
 breaklinks=false,pdfborder={0 0 1},backref=false,colorlinks=false]
 {hyperref}

\makeatletter

\pdfpageheight\paperheight
\pdfpagewidth\paperwidth

\makeatother

\begin{document}

\title{Artificial Intelligence Assists Discovery of Reaction
  Coordinates and Mechanisms from Molecular Dynamics Simulations}


\author{Hendrik Jung}
\thanks{These authors contributed equally to this work.}
\affiliation{Department of Theoretical Biophysics, Max Planck Institute
of Biophysics, 60438 Frankfurt am Main, Germany.}
\author{Roberto Covino}
\thanks{These authors contributed equally to this work.}
\affiliation{Department of Theoretical Biophysics, Max Planck Institute
of Biophysics, 60438 Frankfurt am Main, Germany.}
\author{Gerhard Hummer}
\email[Corresponding author: ]{gerhard.hummer@biophys.mpg.de}
\affiliation{Department of Theoretical Biophysics, Max Planck Institute
of Biophysics, 60438 Frankfurt am Main, Germany.}
\affiliation{Institute of Biophysics,
Goethe University, 60438 Frankfurt am Main, Germany.}

\begin{abstract}
  Exascale computing holds great opportunities for molecular dynamics
  (MD) simulations. However, to take full advantage of the new
  possibilities, we must learn how to focus computational power on the
  discovery of complex molecular mechanisms, and how to extract them
  from enormous amounts of data. Both aspects still rely heavily on
  human experts, which becomes a serious bottleneck when a large
  number of parallel simulations have to be orchestrated to take full
  advantage of the available computing power. Here, we use artificial
  intelligence (AI) both to guide the sampling and to extract the
  relevant mechanistic information. We combine advanced
  sampling schemes with statistical inference, artificial neural
  networks, and deep learning to discover molecular mechanisms from MD
  simulations.  Our framework adaptively and autonomously initializes
  simulations and learns the sampled mechanism, and is thus suitable
  for massively parallel computing architectures.  We propose
  practical solutions to make the neural networks
  interpretable, as illustrated in applications to molecular systems.
\end{abstract}
\maketitle

\section{Introduction}

Computational molecular sciences face two outstanding challenges.  On
the one hand, ever growing computational power and sophisticated
software enable us to simulate increasingly complex systems over ever
longer times, resulting in massive trajectories that we are
ill-equipped to process and interpret. Indeed, we still micromanage
the data production process and rely on visual inspection by human
experts to analyze long simulated trajectories. On the other hand, a
small integration time step and upper bounds to the scalability of
distributed calculations severely limit the sampling of interesting
phenomena in extensive molecular dynamics (MD) simulations. Indeed,
these phenomena are often rare events \cite{Peters2017}, stochastic
transitions between metastable states that are separated by
exponentially longer waiting times. Exascale supercomputers are going to further exacerbate the
interpretation problem and, due to the slowing down of Moore\textquoteright s
law and the \textquotedblleft Communication Wall\textquotedblright ,
fail to fully solve the sampling problem. Two strategies are available to speed up the
sampling \cite{Abrams2013}: use of an unphysical bias in a single MD
simulation to steer the dynamics of the system along a desired
direction; or use of an ensemble of short judiciously initialized
unbiased trajectories to enhance the sampling of transition paths (TPs) between
metastable states. The TPs of rare
events provide invaluable mechanistic insight into the investigated
systems. Often these strategies either completely rely on, or
can greatly profit from, the previous knowledge of a reaction
coordinate (RC) \cite{Rohrdanz2013,Peters2016}, which is a function of
the configuration space that provides a condensed yet accurate
description of the relevant dynamics of the system.

The interpretation and sampling problems are two faces of the same
coin, since learning the salient features of a rare event, its RC,
also enables us to effectively enhance its sampling. The two problems
can therefore be tackled at the same time. However, finding a good RC
is extremely challenging, even for relatively simple dynamical
systems. Moreover, the search for RCs still almost entirely relies on
direct human analysis, through visual inspection and tedious trial and
error approaches, which are potentially biased and frequently fail to
achieve satisfactory results. This dependence on human experts to
initialize and interpret MD simulations threatens to become one of the
main bottlenecks in the exascale computing era. We therefore need to
develop tools able to perform the same tasks with minimal human
intervention.

Trajectories produced by MD simulations are complex and high-dimensional
data, containing subtle patterns that are scarcely identifiable by
human operators, but can be efficiently identified by machine learning
\cite{Bishop2006} and in particular deep learning \cite{Lecun2015,Goodfellow-et-al-2016}
approaches. In fact, specialized deep learning algorithms vastly out-compete
humans in an increasing number of complex data intensive tasks. One
of the challenges, however, is to translate the underlying patterns
into a human understandable form \cite{Lecun2015}.

Here, we show how combining advanced sampling schemes, maximum likelihood
inference, and deep learning allows us to tackle the sampling and interpretation
problems simultaneously, and to move towards the goal of autonomous production
and interpretation of MD simulations of rare events. We take advantage
of the power of Transition Path Sampling (TPS) introduced by Chandler,
Dellago, Bolhuis, Geissler and collaborators \cite{Dellago1998a,Bolhuis2002}
to attempt simulations of transitions between metastable
states. Unbiased simulations started at different initial  
configurations can, at the same time, be used to create a transition
path ensemble \cite{Dellago1998a,Bolhuis2002}, and to estimate the
committor \cite{Hummer2004}, i.e., the probability that
a simulation randomly initialized at a given configuration will evolve
to one metastable state instead another \cite{Peters2016,Peters2017}.
Inspired by the seminal work of Peters and Trout \cite{Peters2006},
we formulate the problem of learning the committor from the outcomes
of the TPS simulation attempts as a statistical inference problem.
Following the pioneering work by Ma and Dinner \cite{Ma2005}, we
use the committor as a mapping between molecular configurations and
the reaction mechanism, that we model by using artificial
neural networks (ANN) \cite{Bishop2006}. By combining all these elements,
we translate the problem of learning molecular mechanisms and RCs
in a quantitative way to the problem of training an ANN on short simulations,
which can be efficiently solved with the methods of deep learning. 

We thus present a novel AI-assisted simulation algorithm that, in
the spirit of reinforcement learning, iteratively and adaptively initializes
and performs new MD simulations, and learns from the generated trajectories
how to increase at every iteration the probability to observe rare
transitions. With minimal human intervention, our algorithm finds
the most efficient way to sample a rare molecular event by learning
the underlying molecular mechanism and the corresponding RC, or, equivalently,
learns the RC of a rare molecular event by sampling it numerous times.
Therefore, we will ultimately obtain not only an unbiased ensemble
of transition trajectories but also the underlying RC encoded in the
trained ANN. 

In the following, we will introduce the main elements of our method
and illustrate its application on a model system. We will then study
a conformational change in the standard molecular benchmark system
alanine dipeptide, and investigate the dissociation of ion pairs in a 
concentred solution of lithium chloride in water.

\section{Algorithm }

\begin{figure}
  \includegraphics[width=1.0\columnwidth]{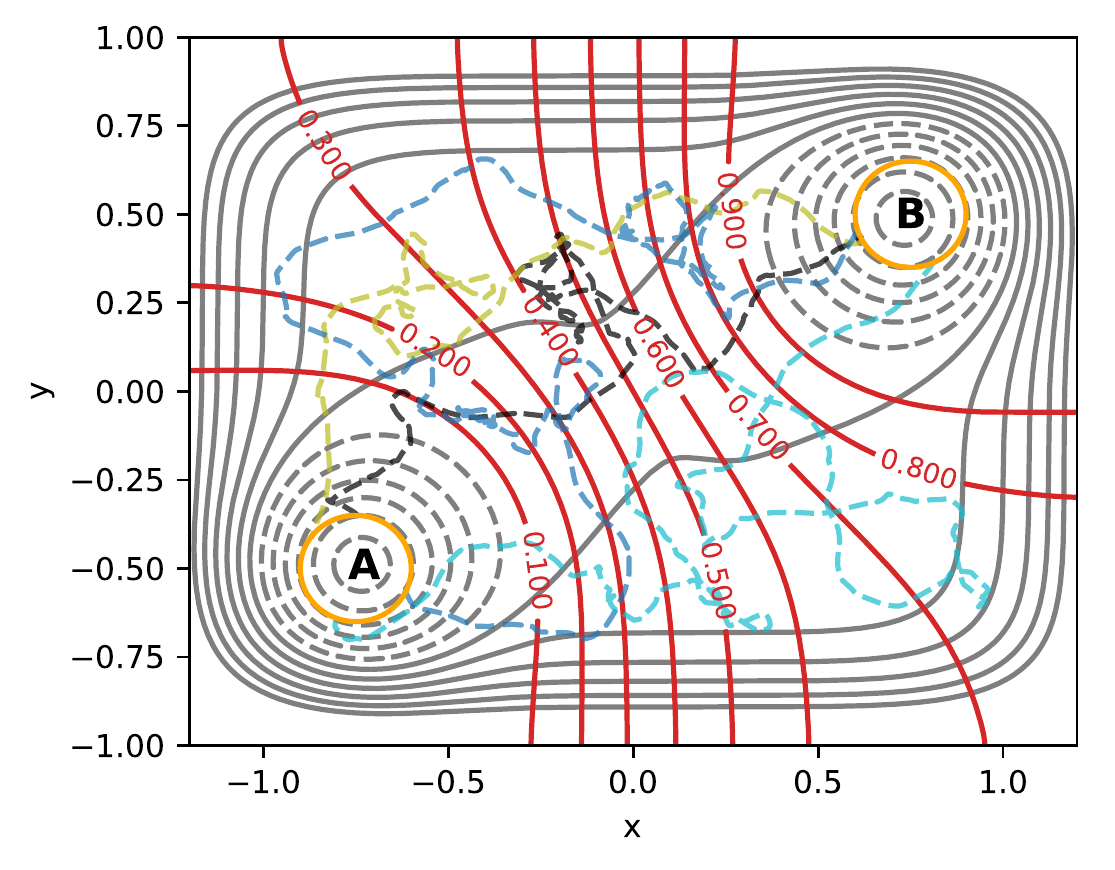}\caption{Model
    potential $V\left(x,y\right)$ (Eq. 1 SI) with two metastable states $A$ and $B$, defined
    as circles of radius $r=0.15$ centered on the minima (orange). 
    The two states are separated by an approximately $7$ $\mathrm{k_{B}T}$ high barrier and a
    broad barrier region. Isolines of the energy surface (grey) are separated
    by $0.7$ $\mathrm{k_{B}T}$.  Isolines of the exact committor
    $p_{B}\left(x,y\right)$ in steps of 0.1 are shown as solid red
    curves.  The dashed colored lines connecting $A$ and $B$ represent
    TPs of varying length simulated by Langevin
    dynamics. \label{fig:Model-potential-}}
\end{figure}

The framework we introduce is valid for transitions among an arbitrary
number of metastable states. However, for the sake of clarity, we will
focus in the following on transitions between two states $A$ and $B$ (Figure
(\ref{fig:Model-potential-})). 

Taking advantage of TPS, we use short unbiased trajectories to
generate an equilibrium ensemble of TPs between $A$ and $B$. Given a path
in configuration space connecting the two states, we initialize MD
simulations from a configuration $\mathbf{x}_{\mathrm{sp}}$ lying on
the path, and iteratively attempt to simulate TPs by using the so
called \textquotedblleft shooting move\textquotedblright{}
\cite{Jung2017}: we redraw velocities $\mathbf{v}$ from the
Maxwell-Boltzmann distribution and start independent simulations from
the configuration $\mathbf{x}_{\mathrm{sp}}$, forward with velocities
$\mathbf{v}$, and backward with velocities $-\mathbf{v}$. Each
simulation is stopped as soon as the trajectory enters one of the two
states.  There can be three possible outcomes: both forward an
backward trajectories reach $A$, or they both reach $B$, or they
reach different states, $A$ and $B$, or $B$ and $A$. In this last case, we
can concatenate the forward and reversed backward trajectories and
form a TP $\chi$. We iterate this sampling scheme by varying the starting
configuration. Each shooting move $i$ is summarized by its starting
configuration, and the number of trajectories $n_{X}^{\left(i\right)}$
ending in states $X=A$ and $B$, i.e., by the set of parameters
$\boldsymbol{\mathbf{\mathbf{\theta}}}^{\left(i\right)}=\left\{
  \mathbf{x}_{\mathrm{sp}}^{\left(i\right)},n_{A}^{\left(i\right)},n_{B}^{\left(i\right)}\right\}
$, with
$n_{A}^{\left(i\right)}+n_{B}^{\left(i\right)}=n^{\left(i\right)}$.
In the following, without loss of generality we will consider
$n^{\left(i\right)}=2$.  We then build the TP ensemble by accepting
simulated TPs based on a Metropolis-Hastings criterion \cite{Jung2017}
(Eq. 2 SI).

The outcome of statistically independent shooting moves follows a
binomial distribution. Therefore, each shooting move is described by
the likelihood function
$L\left[p_{B}(\mathbf{x}_{\mathrm{sp}})\mid n_{A},n_{B}\right]\propto
[p_{B}\left(\mathbf{x}_{\mathrm{sp}}\right)]^{n_{B}}\left[1-p_{B}\left(\mathbf{x}_{\mathrm{sp}}\right)\right]^{n_{A}}$,
where $p_{B}(\mathbf{x}_{\mathrm{sp}})$ and
$p_{A}(\mathbf{x}_{\mathrm{sp}})=1-p_{B}(\mathbf{x}_{\mathrm{sp}})$
are the probabilities that a trajectory initialized at
$\mathbf{x}_{\mathrm{sp}}$ with random velocities will first enter
state $A$ and $B$, respectively.  For statistically independent
trajectories, the quantity $p_{B}(\mathbf{x}_{\mathrm{sp}})$ is known
as the splitting probability or committor of state $B$
\cite{Peters2016}. In our likelihood formulation, we tacitly assume
that the outcome, in terms of the state first reached, for two
trajectories initiated from a common shooting configuration is
independent of the respective initial velocities.

The committor ranges from $p_{B}\left(\mathbf{x}\in A\right)=0$ to $p_{B}\left(\mathbf{x}\in B\right)=1$,
and maps all configurations in the transition region to an intermediate
value (Figure \ref{fig:Model-potential-}). The foliation of level sets
of the committor monitors the progress along the transition $A\rightarrow B$,
and therefore provides a general and rigorous way to quantify the
mechanism described by such a transition \cite{Northrup1982,Du1998,Ma2005,Best2005,Peters2016,Peters2017}.
Consequently, the direction orthogonal to the foliation along the
most probable configurations defines the optimal RC of the transition
$A\rightarrow B$ (Figure \ref{fig:Model-potential-}). Without loss
of generality we model the RC $q\left(\mathbf{x}\right)$ by parametrizing
the committor as 
\begin{equation}
p_{B}\left(\mathbf{x}\right)\equiv p_{B}\left[q\left(\mathbf{x}\right)\right]=\left[1+e^{-q\left(\mathbf{x}\right)}\right]^{-1}.\label{eq:RC}
\end{equation}
The RC is at this point an unknown scalar dimensionless function
$q\left(\mathbf{x}\right)$, which will be in general a complex
nonlinear function of the coordinates of the system. Eq. (\ref{eq:RC})
ensures that the committor ranges between 0 and 1 for any value of the
RC. Also, 
$p_{B}\left[q\left(\mathbf{x}\right)=0\right]=1/2$ defines the
transition state ensemble (TSE) for transitions between two states.

We model the unknown RC $q\left(\mathbf{x}\right)$ with an ANN \cite{Bishop2006}.
ANNs are flexible and powerful tools to model highly nonlinear functions
and reproduce complex patterns in high-dimensional data, and, after
the first pioneering approaches \cite{Ma2005,Behler2007}, were recently
at the center of several important advancements in the field of biomolecular
simulations \cite{Geiger2013,Behler2014,Schneider2017,Sidky,Butler2018,Chen2018,Hernandez2018,Ribeiro2018,Zhang2018,Wehmeyer2018,Mardt2018}.
We thus write the unknown RC as $q\left(\mathbf{x}\right)=q_{\mathrm{ANN}}\left(\mathbf{x}|\mathbf{w}\right)$,
where the weight matrix $\mathbf{w}$, i.e., the fitting parameters
of the ANN, defines the connections between nodes (see Figure 1 and Eq. 3 in the SI for the explicit definition).

After $N$ statistically independent shooting moves, the total likelihood
as a function of the proposed RC will be
\begin{align}
L_{\mathrm{tot}} & \left[q_{\mathrm{ANN}}\left(\mathbf{x}_{\mathrm{sp}}\mid\mathbf{w}\right)\mid n_{A},n_{B}\right]\propto\nonumber \\
 & \prod_{i=1}^{N}\left[p_{B}\left[q_{\mathrm{ANN}}\left(\mathbf{x}_{\mathrm{sp}}^{\left(i\right)}\mid\mathbf{w}\right)\right]\right]^{n_{B}^{\left(i\right)}}\label{eq:Ltot}\\
 & \times\left[1-p_{B}\left[q_{\mathrm{ANN}}\left(\mathbf{x}_{\mathrm{sp}}^{\left(i\right)}\mid\mathbf{w}\right)\right]\right]^{n_{A}^{\left(i\right)}}\nonumber 
\end{align}
Combining equations (\ref{eq:RC}) and (\ref{eq:Ltot}) we obtain
the loss function
\begin{equation}
\begin{aligned}l\left(\mathbf{w}|\boldsymbol{\theta}\right)\equiv-\log L_{\mathrm{tot}}\left[q_{\mathrm{ANN}}\left(\mathbf{x}_{\mathrm{sp}}\mid\mathbf{w}\right)\mid n_{A},n_{B}\right]\\
=\sum_{i=1}^{N}\left[n_{B}^{\left(i\right)}\log\left(1+e^{-q_{\mathrm{ANN}}\left(\mathbf{x}_{\mathrm{sp}}^{\left(i\right)}|\mathbf{w}\right)}\right)\right.\\
+n_{A}^{\left(i\right)}\log\left.\left(1+e^{q_{\mathrm{ANN}}\left(\mathbf{x}_{\mathrm{sp}}^{\left(i\right)}|\mathbf{w}\right)}\right)\right],
\end{aligned}
\label{eq:loss}
\end{equation}
where $\boldsymbol{\theta}=\left\{ \mathbf{x}_{\mathrm{sp}}^{\left(i\right)},n_{A}^{\left(i\right)},n_{B}^{\left(i\right)}\right\} _{i=1,\ldots N}$
is the training set acquired over $N$ shooting moves. We therefore
train the ANN, i.e., we fit the weight matrix $\mathbf{w}$, by minimizing
the loss function Eq. (\ref{eq:loss}) on the training set $\boldsymbol{\theta}$,
and obtain $\hat{q}_{\mathrm{ANN}}\left(\mathbf{x}\mid \mathbf{w}\right)$,
a maximum likelihood estimator of the RC and consequently of the committor.

\begin{figure}
\includegraphics[width=0.65\columnwidth]{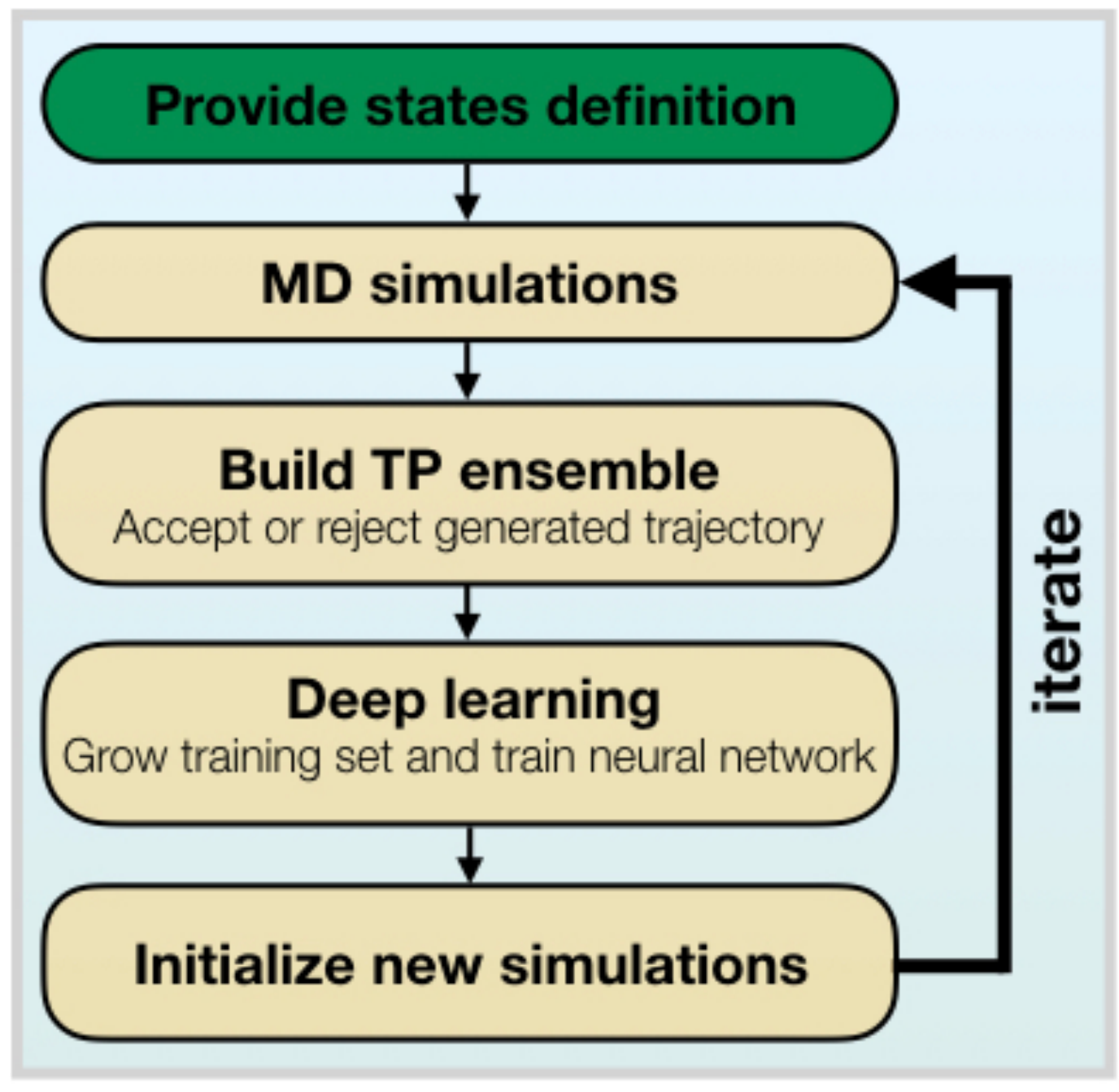}

\caption{\label{fig:flow chart} Schematic flow chart of the AI-assisted MD
simulations algorithm.}
\end{figure}

Recapitulating, the algorithm proceeds in the following way (Figure
\ref{fig:flow chart}). (0) The only required input is the definition
of the metastable states $A$ and $B$ as a function of some order parameters.
We then initialize $q_{\mathrm{ANN}}\left(\mathbf{x}|\mathbf{w}\right)$
by randomly assigning values to the weight matrix $\mathbf{w}$. Given
an initial path $\chi^{\left(0\right)}$ connecting
states $A$ an $B$, we randomly select the first shooting configuration
$\mathbf{x}_{\mathrm{sp}}^{\left(1\right)}$. (1) For every step with
$i>0$ , we perform a shooting move from $\mathbf{x}_{\mathrm{sp}}^{\left(i\right)}$,
generating velocities from a Maxwell-Boltzmann distribution and propagating
the dynamics forward and backward, and append the outcome $\boldsymbol{\theta}^{\left(i\right)}$
to $\boldsymbol{\theta}$. (2) If we sample a new TP, $\chi^{\left(i\right)}$,
we accept or reject it following a Metropolis-Hastings criterion \cite{Jung2017}
(Eq. 2 SI). (3) We minimize the loss function Eq. (\ref{eq:loss})
on the training set $\boldsymbol{\theta}$, which contains all $\boldsymbol{\theta}^{\left(i\right)}$
stored so far, and obtain $\hat{q}_{\mathrm{ANN}}^{\left(i\right)}\left(\mathbf{x}|\mathbf{w}\right)$.
(4) We select a new shooting configuration $\mathbf{x}_{\mathrm{sp}}^{\left(i+1\right)}$
from $\chi^{\left(i\right)}$, the last TP added
to the ensemble, by using the (Cauchy-Lorentz) selection probability
\begin{equation}
p_{\mathrm{sel}}\left(\mathbf{x}|\chi^{\left(i\right)},\mathbf{w},\gamma\right)\propto\frac{\gamma^{2}}{\hat{q}_{\mathrm{ANN}}^{\left(i\right)}\left(\mathbf{x}|\mathbf{w}\right)^{2}+\gamma^{2}},\label{eq:psel}
\end{equation}
which describes a fat-tailed distribution centered around $\hat{q}_{\mathrm{ANN}}^{\left(i\right)}\left(\mathbf{\mathbf{x}^{\mathrm{}}}|\mathbf{w}\right)=0$,
where $\gamma>0$ is the scale parameter. We iterate steps (1) to
(4) until we obtain convergence of the TP ensemble.

The algorithm adaptively creates its own training set, which will
grow at every iteration, and the ANN will be trained both on successes
and failures. The acquired information is then used in Eq. (\ref{eq:psel}),
which ensures that $\mathbf{x}_{\mathrm{sp}}^{\left(i+1\right)}$
is distributed around $\mathbf{x}^{\mathrm{TSE\left(i\right)}}$,
i.e., the best estimate of the TSE given the available training data.
Hence, crucially, at every step the algorithm makes full use of all
available information to autonomously and adaptively initialize new
MD simulations and to optimize the probability to sample new TPs. 

Learning the RC and generating the equilibrium TP ensemble converge
in an orchestrated fashion. During training, the loss function Eq. (\ref{eq:loss})
initially oscillates and then converges, i.e., the ANN converges to
an optimal weight matrix $\mathbf{\mathbf{w^{\mathrm{opt}}}}$ that
can reproduce all future observations $\boldsymbol{\mathbf{\mathbf{\theta}}}^{\left(i\right)}$.
Consequently, at this point we have learned the optimal RC describing
the dynamics of the rare event, encoded in the ANN and defined by
$\hat{q}_{\mathrm{ANN}}^{\mathrm{opt}}\left(\mathbf{x}|\mathbf{w}\right)\equiv\hat{q}_{\mathrm{ANN}}^{\mathrm{}}\left(\mathbf{x}|\mathbf{w^{\mathrm{opt}}}\right)$,
which enables the generation of new TPs with high efficiency. Convergence
of the TP ensemble can be slower and must be monitored separately,
e.g., by tracking the convergence of some physical parameters such as the TP time. As
in similar schemes, if the initial path is very far from the equilibrium
TP ensemble, it might be necessary to discard the TPs sampled during
an initial equilibration phase.

\begin{figure}
\includegraphics[width=1\columnwidth]{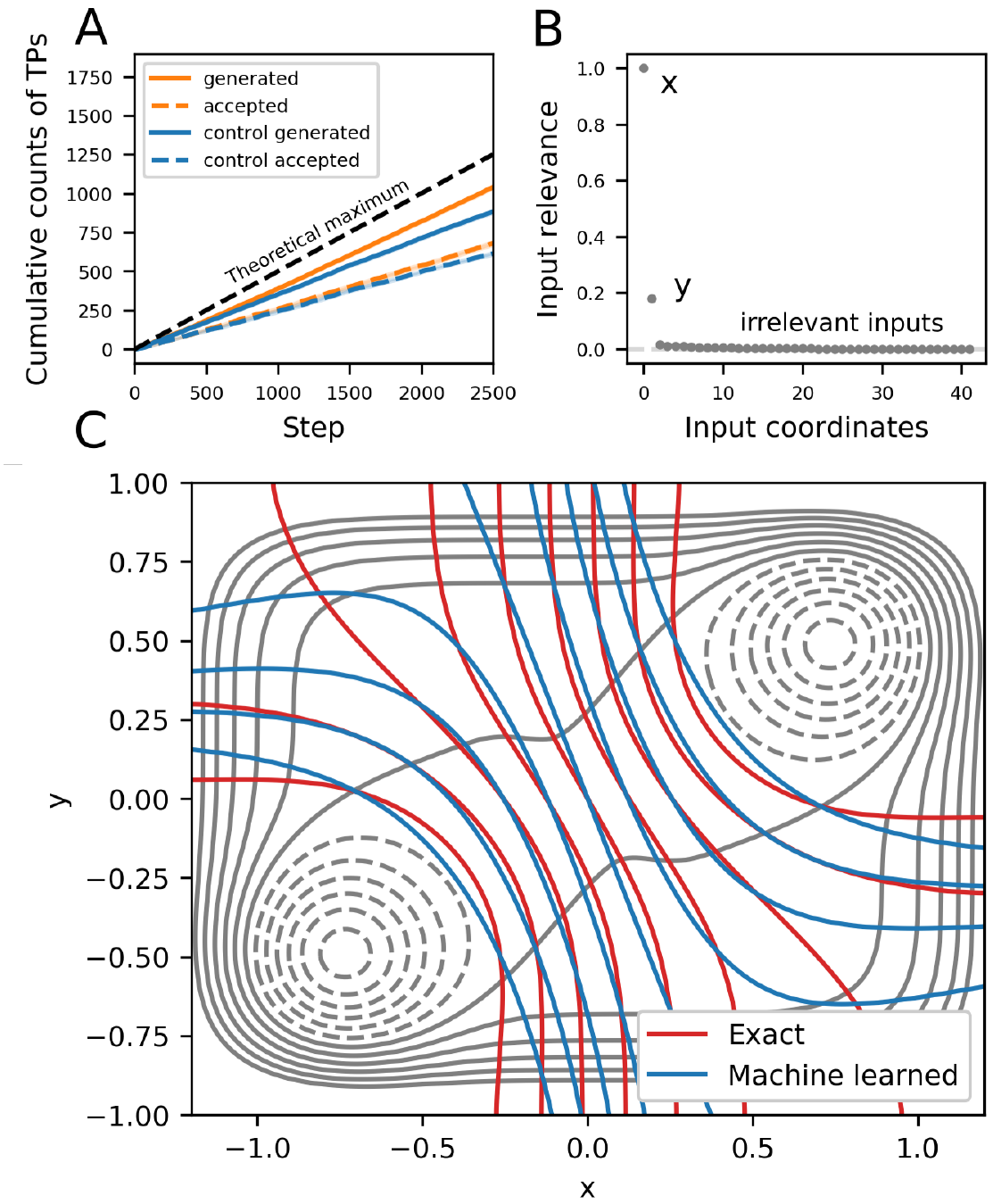}

\caption{Results for 42-dimensional model system of
  Figure~\ref{fig:Model-potential-}.  (A) Cumulative count of
  generated (continuous line) and accepted (dashed line) TPs in
  AI-assisted MD simulations (orange) and standard control TPS (blue)
  performed on the model energy surface shown in Figure
  \ref{fig:Model-potential-}. Each line shows the average over 10
  independent runs. The standard error of the mean is smaller than the
  thickness of the lines. The black dashed line represent the
  theoretical maximum for generation of TPs. (B) Input relevance
  analysis.  All 42 input coordinates are ranked according to their
  normalized relevance. (C) Comparison between the level sets of the
  exact (red) and machine learned (blue) committor
  $p_{B}\left(x,y\right)$.  Isolines are shown in intervals of
  0.1 between 0.1 and 0.9. \label{fig:toymodelefficiencyandHIPRandcommittorcomparison}}
\end{figure}

We illustrate each step of our algorithm by studying a 42-dimensional
model system described by the instantaneous configuration $\mathbf{x}\left(t\right)\in\mathbb{R}^{42}$.
Here, $x_{1}=x$ and $x_{2}=y$ evolve by Langevin dynamics on the
energy potential $V\left(x,y\right)$ shown in Figure \ref{fig:Model-potential-},
which combines the challenges of having two metastable states separated
by a $7$ $\mathrm{k_{B}T}$ high barrier and a broad barrier region.
For each $j>2$, instead, the nuisance coordinate $x_{j}$ evolves
by Langevin dynamics on the harmonic potential $V_{j}\left(x_{j}\right)=\omega_{j}^{2}x_{j}^{2}/2$,
with random angular frequencies $\omega_{j}$. Even though only $x$
and $y$ contribute to the RC, we challenge our deep learning approach
by training the ANN on the full 42-dimensional input space. 

Our algorithm samples TPs with probability close to 50\%, which is the
theoretical upper limit \cite{Hummer2004} assuming statistically
independent shooting moves (Figure \ref{fig:toymodelefficiencyandHIPRandcommittorcomparison}).  Also, it tends to perform better than
the already efficient standard TPS approach, in which initial
configurations $\mathbf{x}_{\mathrm{sp}}^{\left(i+1\right)}$ are
selected with uniform probability from the last accepted TP.

At convergence, our algorithm does not only return the correct equilibrium
TP ensemble but also the RC describing the transition, which is encoded
in $\hat{q}_{\mathrm{ANN}}^{\mathrm{opt}}\left(\mathbf{x}|\mathbf{w}\right)$.
This is a complex function that cannot be immediately interpreted
in a simple way. We can of course always visually inspect the TSE
defined by $\hat{q}_{\mathrm{ANN}}^{\mathrm{opt}}\left(\mathbf{x^{\mathrm{TSE}}}|\mathbf{w}\right)=0,$
which can be trivially extracted from the sampled TPs. Importantly,
the quantitative information contained in the trained ANN can be mapped
on the Cartesian coordinates of molecular structures to provide direct visual insight in the reaction
mechanism. Indeed, we can highlight each atom $n$ according to its
contribution to the RC by evaluating the (normalized) magnitude of
the gradient of the ANN in the TSE, i.e., 
\begin{equation}
c_{n}=\frac{\left|\nabla_{\mathbf{x}^{\left( \mathrm{cart} \right)}_{n}}\hat{q}_{\mathrm{ANN}}^{\mathrm{opt}}\left(\mathbf{x^{\mathrm{TSE}}}|\mathbf{w}\right)\right|}{\left|\sum_{m}\nabla_{\mathbf{x}^{\left( \mathrm{cart} \right)}_{m}}\hat{q}_{\mathrm{ANN}}^{\mathrm{opt}}\left(\mathbf{x^{\mathrm{TSE}}}|\mathbf{w}\right)\right|},\label{eq:contribution to RC}
\end{equation}
which can be calculated by a straightforward application of the chain
rule (backpropagation).

Additionally, we propose a solution to approximately extract the information
encoded in the ANN and translate it to a human friendly explicit formula.
Not all input coordinates $\mathbf{x}$ contribute equally to the
RC. In order to identify the most relevant ones, we build on the procedure
reported in \cite{Kemp2007}. We quantify the information carried
by the $j$-th input coordinate by resampling the loss function (Eq.
(\ref{eq:loss})) replacing one by one each input coordinate $x_{j}$
with random uninformative values. Finally, we rank each input coordinate
$x_{j}$ by the normalized increase in loss caused by their replacement
with noise (Figure \ref{fig:toymodelefficiencyandHIPRandcommittorcomparison}).
This analysis correctly reveals that only two input coordinates actually
determine the output of the ANN, corresponding to $x$ and $y$.

We can now use symbolic regression \cite{Schmidt2009,Izzo2016} to
automatically build a mathematical expression $q_{\mathrm{sr}}\left(x,y\right)$
that approximately reproduces the machine learned optimal RC $\hat{q}_{\mathrm{ANN}}^{\mathrm{opt}}\left(\mathbf{x}|\mathbf{w}\right)$.
Symbolic regression finds the best fit to a given data set searching
both model and parameter space by genetic programming, evolving combinations
of elementary functions and input variables through random mutations
and survival of the fittest \cite{Izzo2016}. Applied to our model
system with input variables $\left(x,y\right)$, symbolic regression
produces a simplified RC, $q_{\mathrm{sr}}\left(x,y\right)=\left(3.32x+1.24y\right)\exp\left(2.92xy\right)$,
which corresponds to a committor in excellent agreement with the exact
one (Figure \ref{fig:toymodelefficiencyandHIPRandcommittorcomparison}).
Indeed, the simplified RC not only captures the linearity of the committor
close to the TSE but also the curvature close to the metastable states.
Clearly, high energy regions of the surface are not sampled by the
MD engine and display a less significant agreement. 

\section{Application to Molecular Systems}

In order to use deep learning to study molecular rare events we must
first decide how to represent molecular systems. Here we face the
trade off between the generality of the representation and finite
sampling. Indeed, it would be possible to use cartesian coordinates
as input for the ANN, but we assume that the systems to simulate will
be challenging, and hence the amount of training data scarce. Therefore,
we will choose general molecular representations that make use of
all previous knowledge already available, e.g., by satisfying the
physical symmetries of the system. For instance, molecular configurations
can be described by using internal coordinates \cite{Parsons2005},
i.e., all bonds, angles, and dihedrals that can be built with a set
of atomic positions. Internal coordinates are in a one-to-one mapping
with cartesian ones, but are manifestly invariant w.r.t. global rotations
and translations. 

A molecular rare event is often modulated by its environment, water in
the simplest case, which should be included in the molecular
representation as well. In the following we will label a ``solute''
the part of the system undergoing the rare transition, and a
``solvent'' its environment. The solvent is invariant under the
permutation of identical particles. We will describe the solvent by
using a modification of symmetry functions introduced by Behler,
Parrinello \cite{Behler2007,Behler2011,Behler2014}, Geiger and Dellago
\cite{Geiger2013}, which are a general way to describe the spatial
organization of molecules around a specific atom.  In particular, we
will use two kind of symmetry functions, $G_{i}^{2}$ and $G_{i}^{5}$
(Eq. (\ref{eq: G2}) and (\ref{eq:G5})). The first quantifies the
density of solvent molecules at a given radial distance from a
specific solute atom; the second quantifies the isotropy of
the angular distribution of solvent molecules at a given radial
distance around a solute atom. These representations are not only
invariant w.r.t. global translations and rotations but also
w.r.t. exchange of solvent molecules.

\subsection{Conformational change in benchmark peptide}

\begin{figure}
\includegraphics[width=1\columnwidth]{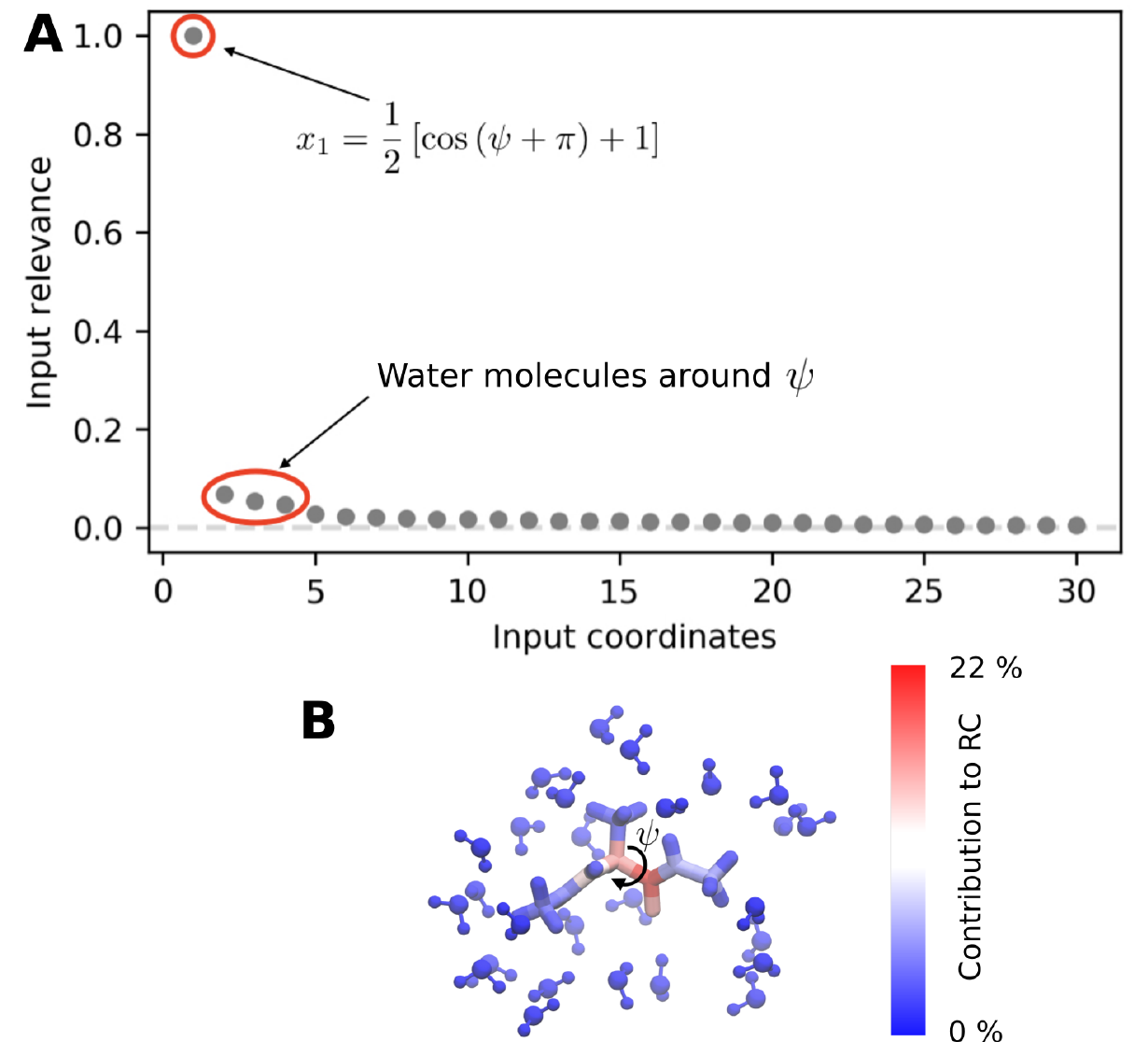}

\caption{\label{fig:ala2}Input relevance analysis and visual representation of
  the machine learned RC for alanine dipeptide. (A) We show the normalized input relevance of
  the 30 most important input coordinates. The first point corresponds
  to the input coordinate
  $x_{1}=\tilde{\psi}\equiv\frac{1}{2}\left[\cos\left(\psi+\pi\right)+1\right]$.
  The following 3 points correspond to coordinates describing the
  organization of water molecules around the peptide. (B)  Structure of alanine dipeptide extracted from a simulated TP that
  is predicted to be part of the TSE. Atomic coordinates are colored
  according to their contribution to the RC (Eq. (\ref{eq:contribution
    to RC}); red most important, blue least important).}
\end{figure}
We applied our computational framework to the prototypical benchmark
peptide alanine dipeptide ($N$-acetyl alanine $N'$-methylamide)
\cite{Bolhuis2000}. We investigated the thermally activated conformational
change of the dihedral angle $\psi$, corresponding to trajectories
crossing a large free energy barrier as shown in Figure 2 of the SI.
The only required initial input are the metastable states, defined
in this case as a function of the dihedral angles $\phi$ and $\psi$.
We simulated the system in explicit solvent, and used internal coordinates
to describe the peptide configurations and symmetry functions to represent
the solvent, using both as inputs for deep learning. Our framework enabled us to simulate TPs with high efficiency, and
rapidly obtain an equilibrated TP ensemble (Figure 3 SI). 

The input relevance analysis reported in Figure \ref{fig:ala2} correctly
identifies the coordinate
$x_{1}=\tilde{\psi}\equiv\frac{1}{2}\left[\cos\left(\psi+\pi\right)+1\right]$
as the major descriptor of the conformational change. The analysis
also highlights a marginal role for three further input coordinates
describing the organization of water molecules around the peptide.  In
particular, these three coordinates describe the angular distribution
of water oxygens at a distance of $0.175\,\mathrm{nm}$ from the oxygen
of the acetyl involved in the peptide bond ($x_{2}$), a hydrogen atom
in the methyl-group ($x_{3}$), and the nitrogen of the methylamide
($x_{4}$) (Table I SI). Figure \ref{fig:ala2} shows a representative
structure of the TSE, for which the ANN predicts a value of
$p_{B}=0.5$ in excellent agreement with the $p_{B}=0.46$ value returned
by a direct sampling of the committor using 10,000
trajectories. Each atom in the structure is colored according to its
contribution to the RC, and a simple visual inspection immediately
reveals the dominant aspect of the molecular mechanism.

We then used the four most relevant input coordinates to approximate
the trained ANN by a simple function. Depending on the desired level
of regularization we obtained slightly different expressions (Table II
SI). The optimal is $q_{\mathrm{sr}}\left(\tilde{\psi},x_{3}\right)=-8.38\tilde{\psi}+5.62\exp\left(-0.285x_{3}\right)$,
which evaluated on the training data returns a loss of $1.051$ compared
to $1.031$ for the full ANN.

\subsection{Ion dissociation in water}

\begin{figure}
  \includegraphics[width=1\columnwidth]{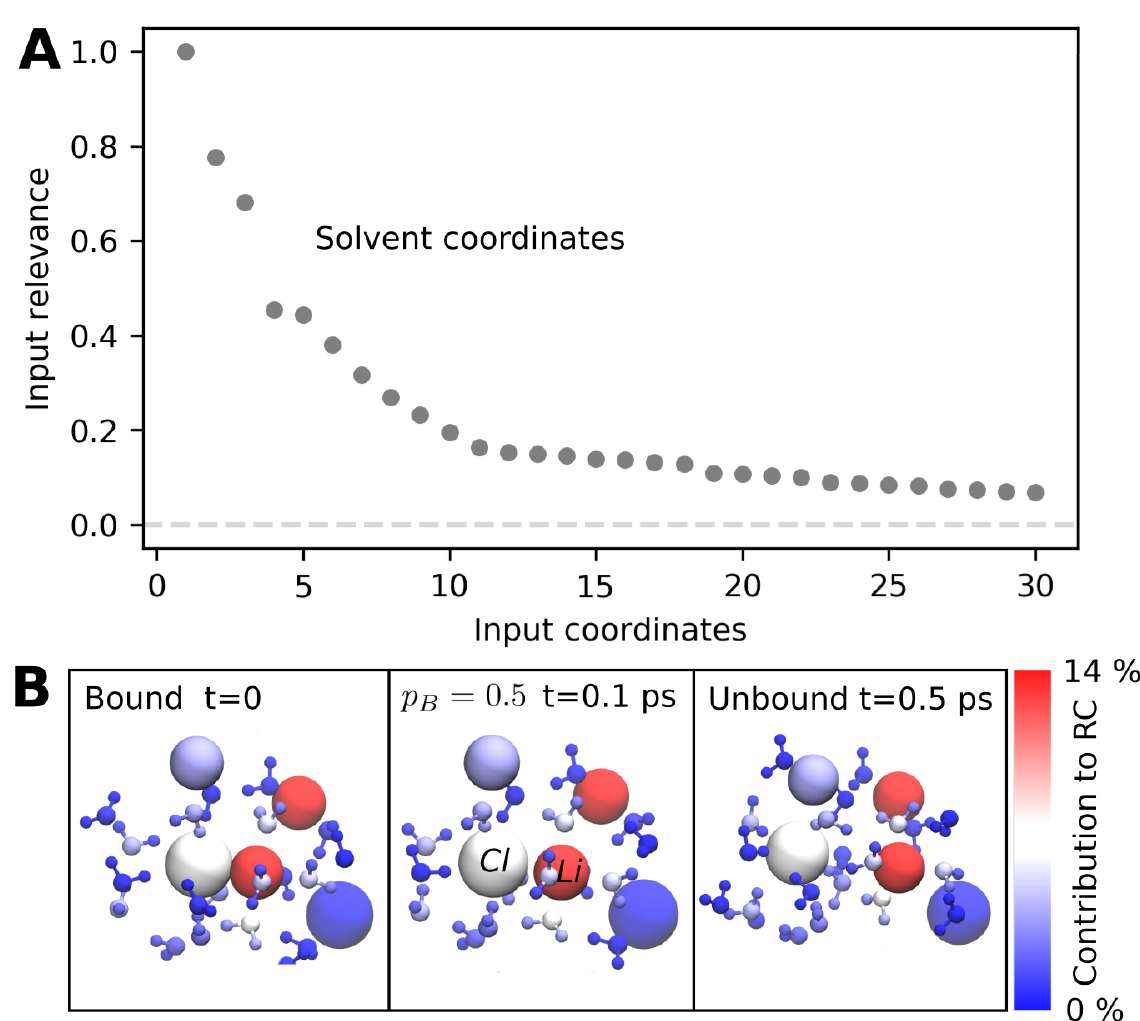}
  \caption{\label{fig:iondissociation} Input relevance analysis and visual representation of the machine
    learned RC for LiCl dissociation. 
    (A) We report the normalized input relevance of the 30
    most important input coordinates, which all describe the
    collective distribution of the solvent around the dissociating ion
    pair. (B)  Configurations extracted from a simulated
    TP, in the bound state, from the TSE, and in the unbound state. 
    The dissociating ion
    pair is shown at the center and labeled (small spheres: Li$^+$;
    large spheres: Cl$^{-}$). Atomic coordinates are colored according
      to their contribution to the RC (Eq. (\ref{eq:contribution to
        RC}); red most important, blue least important).}
\end{figure}

We then investigated the dissociation of lithium and chloride ions
in a $1\,\mathrm{M}$ solution. Despite its apparent simplicity, understanding
this reaction in quantitative terms poses a challenge \cite{Geissler1999},
because it is governed by many-body long range interactions and by
instantaneous collective motions of all surrounding water molecules
and ions. We focused on a specific pair of ions, and used their distance
$r$ as a parameter to define bound and unbound states. We then described
the remaining part of the solution with symmetry functions of the
positions of all other lithium and chloride ions, and the oxygen of
water molecules. Our AI-assisted simulation
framework produces TPs significantly more efficiently than a standard TPS implementation
(Figure 4 SI).

The input importance analysis reveals that indeed the environment
controls the dissociation reaction (Figure \ref{fig:iondissociation}
and Table III SI). Contrary to the previous application, there is
no single dominant descriptor of the reaction. The trained ANN employs
a large number of input coordinates, and their importance decays in
a power-law fashion. Of the 10 most relevant coordinates, most
probe the density and angular distribution of lithium and chloride
counter-ions around the dissociating ion pair, while only one probes
the distribution of water molecules (Table III SI). Notably, in independent runs
we obtain alternative ``mirror'' symmetry functions, where lithium
and chloride exchange, which means that the ANN can use alternative
descriptions of analogous configurations. Also, the interatomic distance
of the dissociating pair never appears among the most important input
coordinates. Figure \ref{fig:iondissociation} shows representative configurations along a TP, including a
configuration of the TSE identified by the trained network, for which
a direct sampling of the committor with 1,000 simulations returns
$p_{B}=0.53$. The structures are colored according to the contribution
of each atom to the RC, and immediately reveal the many-body character
of the mechanism and its main features. Indeed, we can clearly see
4 water molecules coordinating the central lithium ion and 2 distant
competing lithium counter-ions.

Distilling a simple equation from the full network is in this case
more challenging. By using the 10 most relevant input coordinates
identified in Figure \ref{fig:iondissociation}, we obtained expressions
that do not correctly reproduce the committor of the reaction. This
is possibly due to an insufficient number of input coordinates. We obtained
accurate expressions by using the first 10 coordinates and the interatomic
distance $r$ of the dissociating ion pair. Depending on the regularization,
we obtained a number of different expressions that all offer an insightful
physical interpretation (Table IV SI). As an example, we report
$q_{\mathrm{sr}}\left(r,x_{1},x_{9}\right)=29.5r-8.19\exp(-0.87x_{1})-0.29x_{9}$,
where $x_{1}$ describes the angular distribution of distant ($\sim0.7$
nm) chloride counter-ions and $x_{9}$ describes the angular distribution
of close ($\sim0.175$ nm) water molecules around the dissociating
lithium ion.

\section{Discussion}

The computational framework we introduced enhances the simulation
of rare events with little previous knowledge and minimal human intervention.
Given two metastable states defined with some arbitrary order parameters,
the algorithm efficiently simulates rare events by learning the underlying
reaction mechanism, using only adaptively initialized short and unbiased
simulations. In the spirit of reinforcement learning \cite{Bishop2006},
our framework autonomously builds its own training set, training both
on successes and failures. We combine a number of established techniques
in a very general way, such that each building block might be exchanged
with a potentially more suitable alternative. This could involve the
sampling strategy \cite{Grunwald2008,Mullen2015,Menzl2016}; performing
MD simulations at atomistic or coarse-grained resolution; considering
more sophisticated inference models and integrating experimental information
in a systematic way; exploiting alternative network architectures
and different regression schemes for the learning phase \cite{Mones2016}. 

At convergence, the algorithm returns an equilibrium ensemble of simulated
TPs and a functional form of the committor encoded in an ANN, which
is both differentiable and inexpensive to evaluate. We therefore obtain
a very efficient way to calculate the committor of molecular configurations,
which can be used to identify level sets of the committor from the
sampled TPs by straightforward evaluation and bookkeeping. Given that
these structures belong to unbiased trajectories guarantees that the
committor is evaluated only in the relevant regions of the configuration
space. Molecular configurations belonging to the level sets can then
immediately be used to further characterize the transition with methods
like milestoning \cite{Faradjian2004}, forward-flux-sampling \cite{Allen2009},
and s-shooting \cite{Menzl2016}.

Despite its crucial role in controlling molecular mechanisms, we often
neglect the environment in which a rare event takes place, mostly
because we struggle to capture many-body interactions. By processing
large amounts of input coordinates, machine learning offers the potential
to systematically describe even complex biomolecular environments,
like dense solutions or complex lipid bilayers. 

We proposed a solution to make the trained network interpretable,
by approximating the ANN with a simple function of the few input coordinates
that actually control the transition. The resulting equation is an
accurate approximation of the RC that can be used to calculate the
free energy surface governing the transition and its kinetics by using
techniques like umbrella sampling \cite{Kastner2011,grossfield_wham},
metadynamics \cite{Parrinello2002}, and within the framework of TPS
itself \cite{Best2005}. The analytic form of the
RC also provides a starting point to inspire and build physical analytic theories.
Any previous knowledge can be easily integrated with the generated
equation, simply by imposing any relevant coordinate or functional
form that one wants to appear in the reduced RC representation. Additionally,
visualizing the contribution to the RC with automatically generated
annotated structures is a powerful tool to gain insight in the reaction
mechanism (Figure \ref{fig:ala2} and \ref{fig:iondissociation}).
The AI autonomously generates images guiding human operators in
understanding the discovered mechanism, making the ANN interpretable
in a very immediate way.

The resampling procedure to identify the important input coordinates
is particularly effective when those have a physical meaning. However,
for challenging problems we will usually use coordinates that are
agnostic and redundant. We expect that in these cases the resampling
procedure will return a power-law spectrum, similar to what we observe
in Figure \ref{fig:iondissociation}, and extracting few essential
input coordinates will be challenging. This highlights the importance
of developing representations of given classes of systems
general enough to be transferable within the same class \cite{Bartok2013,Bartok2017}.
Similarly to other recent strategies, we can easily feed a large list
of features (``collective variables'') to our algorithm \cite{Chen2015}.
Another route to follow is to combine deep learning and similar methods
with manifold learning techniques \cite{Chen2018}. 

Future versions of the algorithm will benefit from the ongoing research
in the field of interpretable ANN. In particular, we will explore
the possibility that networks trained on molecular events learn a
hierarchy of features of the systems, similarly to what happens with
hierarchical networks used in the field of image processing. One strategy
is to map back the contribution of each layer of a deep network to
molecular structures. These hierarchical features could be used to
engineer transferable representations used as inputs for other problems,
as discussed above. 

An optimal definition of the metastable states is an important aspect
to avoid wasting computational resources, and can be improved within
the present framework in an iterative way. One can start with a very
conservative definition, and then merge regions of the configuration space
that the network, trained on quick exploratory runs, estimates being strongly
committed to either one of the states. 

The algorithm can be easily generalized to study rare transitions
between multiple states. This is straightforward if the states
are known in advance, since it suffices to consider a multi-state
TPS \cite{Rogal2008} and definition of the committor \cite{Kiraly2018}
and to use a multinomial equivalent of Eq. (\ref{eq:RC}), i.e., a
softmax function. With the same approach we can treat trajectories
that will not commit during the available simulation time because
of unknown states or long-lived intermediates, which can be identified
and characterized in post-processing. Simulations that do not commit
because of extremely long TP times, instead, will require a staged
approach and an opportune modification of the inference model.

Controlling the balance between exploitation and exploration is
straightforward in our framework. Maximum efficiency of generating TPs
(exploitation) is obtained when the network initializes new
trajectories close to the TSE, which occurs when the selection
probability in Eq.  (\ref{eq:psel}) is sharply peaked about the
TSE. An ideal RC, however, should explain the transition not only
close to the TSE but also in all other regions of the reaction, and in
particular close to the boundaries of the metastable
states. Exploration can thus be prioritized by a flatter selection
probability that increases the probability to initialize trajectories
far from the TSE.

Tuning the selection probability to initialize trajectories away from
the TS, and also within the metastable states, facilitates the
discovery of additional reaction channels and therefore alternative
mechanisms within the same exploratory run.  Alternatively, one could
train two or more independent networks in parallel. Comparing the
losses evaluated by the first network on configurations sampled by the
second one \textendash{} and vice versa \textendash{} will reveal
whether the two networks have indeed sampled the same
mechanism. Furthermore, we will explore more systematic ways of
discovering multiple mechanisms by using multi-domains networks and
inference schemes \cite{Wu2018}.

Our algorithm shares with other data-driven learning approaches the
challenges of avoiding overfitting and of validating the learned
results. The committor, which is ultimately what the ANN will learn,
can always be validated in an expensive yet straightforward direct
sampling. This will not be an option, however, for systems that are
extremely expensive to simulate, for which more advanced strategies
are required, like designing novel training schemes that intervene
only when new ``unusual'' configurations are sampled, and using
ANN with built-in regularizations. However, overfitting is not a serious issue for the production of data,
and could be avoided in a thorough post-processing to learn the RC.

Regarding scalability the algorithm is already almost trivially parallelizable.
Indeed, all the deep learning machinery can be concentrated on a single
master node, which at every iteration communicates initial configurations
to other nodes in the network, harvests the outcome of the performed
simulations, and trains the artificial neural network on the resulting
new data. Since every step of this distributed calculation satisfies
detailed balance, the resulting sets of simulated TPs can simply be
combined to obtain an equilibrium ensemble. This feature potentially
solves the challenge of efficiently scaling simulations of rare events
on large supercomputers.

In conclusion, we introduced a novel AI-assisted computational
framework that represents a step forward towards the goal of automatic
design, execution, and interpretation of MD simulations of rare
events. This framework simultaneously enables the enhancement of
sampling, the discovery of reaction mechanism, and the learning of the
corresponding reaction coordinate, thereby opening the door to
simulations of complex molecular events that are currently
unfeasible. Our algorithm has the potential to transform MD
simulations of rare and interesting molecular events into a
semi-autonomous high-throughput technique, capable of scaling
efficiently on exascale supercomputers. Using reinforcement learning
to learn the committor is a very general idea, which can be
potentially used to study any time-series of complex data that admits
a forward commitment probability. The main challenge for the future
will be to make the ANN systematically interpretable and distilling
analytic representations \textendash{} or even physical laws
\cite{Schmidt2009} \textendash{} that describe the discovered
molecular mechanism in a more human-accessible form.

\section{Materials and Methods}

\subsection{Model system}

We simulated Langevin dynamics on the model energy landscape shown
in Figure \ref{fig:Model-potential-} described by Eq. 1 SI using the openpathsampling engine
\cite{Swenson2018} and an integrator with BAOAB splitting \cite{Leimkuhler2013},
with a timestep $\Delta t=0.02~\mathrm{ns}$ 
and friction $\gamma=2.5~\mathrm{ns^{-1}}$. We defined the states
$A$ and $B$ as circles with a radius of $r=0.15$ centered
on the two minima. Each additional independent nuisance coordinate
$x_{i}$ follows a harmonic potential with frequency $\omega_{i}$.
The first two frequencies were chosen such that their periods are
respectively larger and comparable to the average transition path
time of $\langle t_{TP}\rangle \approx 11~\mathrm{ns}$. The remaining frequencies
were chosen randomly from an uniform distribution in the interval
$[0~\mathrm{ns^{-1}},10~\mathrm{ns^{-1}})$. All oscillator masses
were set to $m=1$. The initial TP is a straight line connecting the two
stable states in the plane of the coordinates $x$ and $y$. The exact
committor reported in Figures \ref{fig:Model-potential-} and \ref{fig:toymodelefficiencyandHIPRandcommittorcomparison}
was obtained by solving the adjoint Fokker-Planck equation with a
standard relaxation technique. 

\subsection{Deep learning}

We performed deep learning by training a self-normalizing ANN
\cite{Klambauer2017} with four hidden layers and a final layer with one node and
a linear activation function. We used the \textquotedbl{}Lecun normal\textquotedbl{}
initialization and applied a dropout of $10\%$ after the last hidden
layer \cite{Klambauer2017}. We trained the ANN with the Adam gradient
descent algorithm \cite{Kingma2014}, training only after every second
TPS step and using a learning rate of $\mathrm{lr}=5\cdot10^{-4}$.
We interrupted the training process if the loss stayed almost constant
and resumed it only upon a loss decrease. For the model system the
number of nodes per layers equals the number of inputs, 42. For simulations
of alanine dipeptide and LiCl dissociation it equals half the number
of inputs, hence 750 and 264 nodes per layer, respectively. We performed
all deep learning with custom written code based on keras \cite{chollet2015keras}. 

\subsection{Input coordinate relevance analysis}

For every $j$-th input coordinate we built the set $\tilde{\boldsymbol{\theta}}\left(j\right)=\left\{ \tilde{\mathbf{x}}_{\mathrm{sp}}^{\left(i\right)}\left(j\right),n_{A}^{\left(i\right)},n_{B}^{\left(i\right)}\right\} _{i=1,\ldots N}$,
where $\tilde{\mathbf{x}}_{\mathrm{sp}}^{\left(i\right)}\left(j\right)$
was obtained by replacing the $j$-th component of $\mathbf{x}_{\mathrm{sp}}^{\left(i\right)}$
with $\tilde{x}_{j}^{\left(i\right)}\sim\mathcal{U}\left(\min\left(x_{j,\mathrm{sp}}\right),\max\left(x_{j,\mathrm{sp}}\right)\right)$,
i.e., a random number uniformly distributed between the extreme values
of $x_{j,\mathrm{sp}}$ over the whole training set $\boldsymbol{\theta}$.
We then defined the input relevance $r\left(x_{j}\right)=\left[l\left(\mathbf{w}^{\mathrm{opt}}|\tilde{\boldsymbol{\theta}}\left(j\right)\right)-l\left(\mathbf{w}^{\mathrm{opt}}|\boldsymbol{\theta}\right)\right]$,
which quantifies the impact of the lost information carried by the
$j$-th input coordinate by resampling the loss function (\ref{eq:loss}).
This quantity is larger the more informative $x_{j}$ is as an input
coordinate. We normalized the input relevance by subtracting the converged
loss of the unperturbed ANN and dividing all entries by the largest
relevance. We then ranked each input coordinate $x_{j}$ by the normalized
$r\left(x_{j}\right)$ and identified the most relevant ones. 

\subsection{Symbolic regression}

The implementation of the symbolic regression is based on the python
package dcgpy \cite{Izzo2016}, and alternates between genetic programming
optimization and gradient based optimization of the constants. The
fitness of each trial expression $q_{\mathrm{sr}}\left(\mathbf{z}\right)$
is measured by $l_{\mathrm{sr}}\left(q_{\mathrm{sr}}|\boldsymbol{\theta}\right)\equiv-\log L_{\mathrm{tot}}\left[q_{\mathrm{sr}}\left(\mathbf{z}_{\mathrm{sp}}\right)\right]+\lambda n_{c}$,
where $\mathbf{z}$ is a subset of all input coordinates $\mathbf{x}.$
We added the regularization term $\lambda n_{c}$ in order to keep
expressions simple and avoid overfitting, with $\lambda>0$ and $n_{c}$
being a measure of the complexity of the trial expression. 

\subsection{Molecular representation}

In molecular systems one often distinguishes a solute and a solvent. We described
configurations of the first by using internal coordinate \cite{Parsons2005},
and the second by using a modification of previously introduced symmetry
functions \cite{Behler2007,Behler2011,Geiger2013,Behler2014}. The
first one quantifies the density of solvent molecules around a solute
atom $i$ and is defined as:
\begin{equation}
G_{i}^{2}=\sum_{j}e^{-\eta\left(r_{ij}-r_{s}\right)^{2}}f_{c}\left(r_{ij}\right),\label{eq: G2}
\end{equation}
where the sum runs over all solvent atoms $j$ of a specific atom
type, $r_{ij}$ is the distance between the central atom $i$ and
atom $j$, $r_{s}$ is the distance from the central atom at which
the shell is centered and $\eta$ controls the width of the shell.
The function $f_{c}(r)$ is a Fermi cutoff defined as:
\begin{equation}
f_{c}\left(r\right)=\begin{cases}
\left[1+\exp\left(\alpha_{c}\left(r-r_{c}-1/\sqrt{\alpha_{c}}\right)\right)\right]^{-1} & r\leq r_{c}\\
0 & r>r_{c}
\end{cases},
\end{equation}
which ensures that the contribution of distant solvent atoms vanishes.
The second type of symmetry function probes the angular distribution
of the solvent in a shell around the central atom $i$:
\begin{alignat}{1}
G_{i}^{5}=\sum_{j,k>j}\left[\lambda+\cos\vartheta_{ijk}\right]^{\zeta} & e^{-\eta\left[\left(r_{ij}-r_{s}\right)^{2}+\left(r_{ik}-r_{s}\right)^{2}\right]}\times\label{eq:G5}\\
\times f_{c}\left(r_{ik}\right)f_{c}\left(r_{ij}\right),\nonumber 
\end{alignat}
where the sum runs over all solvent atom pairs, $\vartheta_{ijk}$ is
the angle spanned between the two solvent atoms and the central solute
atom, the parameter $\zeta$ must be an even number and controls the
sharpness of the angular distribution, while the parameter $\lambda$
is either +1 or \textminus 1 and controls whether the location of
the minimum of the cosine bracket is at $\vartheta_{ijk}=0$ or at $\vartheta_{ijk}=\pi$.
Training the ANN is facilitated if all inputs lie in a range between
0 and 1. We therefore used $0.5(\cos(\phi)+1)$ for every angle $\phi$,
and $\left(0.5(\sin(\vartheta)+1),~0.5(\cos(\vartheta)+1)\right)$ for every
dihedral angle $\vartheta$. We normalized the symmetry functions, dividing
by the expected average number of atoms or pairs for an isotropic
distribution in the probing volume.

\subsection{MD simulations}

We simulated alanine dipeptide in the amber ff99SB-ILDN forcefield
\cite{AMBER99SB-ILDN}, with the termini capped by $N$-acetyl and
$N$-methylamide groups, and solvated by 543 TIP3P water molecules
\cite{Jorgensen1983}.

We simulated a solution of lithium chloride in the Joung and Cheatham
forcefield \cite{Joung2008}, using a cubic simulation box containing
37 lithium and 37 chloride ions, solvated with 2104 TIP3P water molecules,
corresponding to a concentration of 1 M LiCl.

We performed all MD simulations using the openMM MD engine \cite{Eastman2017},
with a velocity Verlet integrator with velocity randomization \cite{Sivak2014}
from the python package openmmtools, using a time step of $\Delta t=2~\mathrm{fs}$.
All simulations are performed in the NVT-Ensemble, at a temperature
of $T=300~\mathrm{K}$. The friction is set to $\gamma=1/\mathrm{ps}$,
non-bonded interactions are calculated using a particle mesh Ewald
scheme \cite{Essmann1995} with a cutoff of 1 nm and an error tolerance
of $0.0005$. We constrained the length of all bonds involving hydrogen
atoms. 

\subsection{TPS}

All TPS simulations were carried out using a customized version
of the openpathsampling python package \cite{Swenson2018} together
with custom python code. For alanine dipeptide the states $A$ and $B$
are defined in terms of the dihedral angles $\psi$ and $\phi$. For
both states $\phi$ must lie in the range $[-\pi,0]$. For state $A$
$\psi$ must lie in the range $\left[\frac{2}{3}\pi,\frac{10}{9}\pi\right]$,
while for state $B$ $\psi\in\left[\frac{5}{18}\pi,\frac{1}{6}\pi\right]$.
We consider LiCl ions bound if $r\le0.24~\mathrm{nm}$, and as unbound
if $r\ge0.47~\mathrm{nm}.$ We obtained initial TPs by standard MD
at a temperature of $T=400~K$.  New shooting configurations $\mathbf{x}_{\mathrm{sp}}$
were selected from the last accepted TP with probability reported
in Eq. (\ref{eq:psel}) with $\gamma=1$. 

\medskip{}

We used custom code written using NumPy \cite{Oliphant}, SciPy \cite{Jones},
IPython \cite{Perez2007}, Matplotlib \cite{Hunter2007}, Cython \cite{Behnel2011},
SymPy \cite{Meurer2017}, and mdtraj \cite{McGibbon2015} to analyze
and visualize data. Molecular representations were made with VMD \cite{Humphrey1996}.

\medskip{}
\begin{acknowledgments}
The authors acknowledge Dr. Anna Kreshuk, Profs. Peter Bolhuis, Christoph
Dellago, Andrew Ferguson, and Mark E. Tuckerman, for stimulating discussions,
and the Open Path Sample community, in particular Dr. David Swenson,
for discussions and technical support. This work was supported by
the Max Planck Society. 
\end{acknowledgments}

\bibliographystyle{apsrev4-1}
\bibliography{biblio}

\end{document}


\title{Supporting Information: Artificial Intelligence Assists Discovery of Reaction
  Coordinates and Mechanisms from Molecular Dynamics Simulations}

\author{Hendrik Jung}
\thanks{These authors contributed equally to this work.}
\affiliation{Department of Theoretical Biophysics, Max Planck Institute
of Biophysics, 60438 Frankfurt am Main, Germany.}
\author{Roberto Covino}
\thanks{These authors contributed equally to this work.}
\affiliation{Department of Theoretical Biophysics, Max Planck Institute
of Biophysics, 60438 Frankfurt am Main, Germany.}
\author{Gerhard Hummer}
\email[Corresponding author: ]{gerhard.hummer@biophys.mpg.de}
\affiliation{Department of Theoretical Biophysics, Max Planck Institute
of Biophysics, 60438 Frankfurt am Main, Germany.}
\affiliation{Institute of Biophysics,
Goethe University, 60438 Frankfurt am Main, Germany.}

\begin{abstract}
  Exascale computing holds great opportunities for molecular dynamics
  (MD) simulations. However, to take full advantage of the new
  possibilities, we must learn how to focus computational power on the
  discovery of complex molecular mechanisms, and how to extract them
  from enormous amounts of data. Both aspects still rely heavily on
  human experts, which becomes a serious bottleneck when a large
  number of parallel simulations have to be orchestrated to take full
  advantage of the available computing power. Here, we use artificial
  intelligence (AI) both to guide the sampling and to extract the
  relevant mechanistic information. We combine advanced
  sampling schemes with statistical inference, artificial neural
  networks, and deep learning to discover molecular mechanisms from MD
  simulations.  Our framework adaptively and autonomously initializes
  simulations and learns the sampled mechanism, and is thus suitable
  for massively parallel computing architectures.  We propose
  practical solutions to make the neural networks
  interpretable, as illustrated in applications to molecular systems.
\end{abstract}
\maketitle

\section{Model system}

The functional form of the model energy surface shown in Fig. 1
is
\begin{align}
V(x,y) & =  \sigma_{x}x^{6}+\sigma_{y}y^{6} \nonumber \\
 & +  \epsilon_{A}\exp\left(-\alpha_{x}^{A}(x-x_{0}^{A})^{2}-\alpha_{y}^{A}(y-y_{0}^{A})^{2}\right)\label{eq:model surface}\\
 & +  \epsilon_{B}\exp\left(-\alpha_{x}^{B}(x-x_{0}^{B})^{2}-\alpha_{y}^{B}(y-y_{0}^{B})^{2}\right)\nonumber 
\end{align}
where $\sigma_{x}$ and $\sigma_{y}$ are parameters controlling the
steepness of the outer boundary, and $\epsilon_{A}$ and $\epsilon_{B}$
control the depth of state $A$ and $B$, respectively. The states have
minima located at positions $\left(x_{0}^{A},y_{0}^{A}\right)$ and
$\left(x_{0}^{B},y_{0}^{B}\right)$, with extensions controlled by
$\left(\alpha_{x}^{A},\alpha_{y}^{A}\right)$ and $\left(\alpha_{x}^{B},\alpha_{y}^{B}\right)$. 
We expressed energies in inverse temperature units, $\beta=\left( \mathrm{k_B T}\right)^{-1}$, with $\mathrm{k_B}$ the Boltzmann constant and $T$ the absolute temperature. 
We performed simulations with constants $\sigma_{x}=2~\mathrm{\beta}^{-1}$
and $\sigma_{y}=10~\mathrm{\beta}^{-1}$, and $\alpha_{x}^{A}=\alpha_{y}^{A}=\alpha_{x}^{B}=\alpha_{y}^{B}=12$.
The barrier height was set to approximately $7\beta^{-1}$ by
choosing $\epsilon_{A}=\epsilon_{B}=-7~\mathrm{\beta}^{-1}$. The states
$A$ and $B$ are located at $\left(x_{0}^{A}=-0.75,y_{0}^{A}=-0.5\right)$
and $\left(x_{0}^{B}=0.75,y_{0}^{B}=0.5\right)$.

\section{Transition path sampling}

If transition path (TP) shooting produces a new TP $\chi^{\left(i\right)}$,
we accept or reject it with probability $p_{\mathrm{acc}}$ given by the  Metropolis-Hastings
criterion \cite{Jung2017}.
\begin{equation}
p_{\mathrm{acc}}(\chi^{\left(i\right)}\mid\chi^{\left(i-1\right)}\left(\tau^{\prime}\right))=\min\left(1,\frac{p_{\mathrm{sel}}(\mathbf{x}_{\mathrm{sp}}\mid\chi^{\left(i\right)})}{p_{\mathrm{sel}}(\mathbf{x}_{\mathrm{sp}}\mid\chi^{\left(i-1\right)}\left(\tau^{\prime}\right)}\right),\label{eq:MH acceptance}
\end{equation}
where $p_{\mathrm{sel}}(\mathbf{x}_{\mathrm{sp}}\mid\chi^{\left(i\right)})$
is the probability to select a particular shooting configuration from
the TP. 

\section{Deep Learning}

\begin{figure}
  \includegraphics[width=0.70\columnwidth]{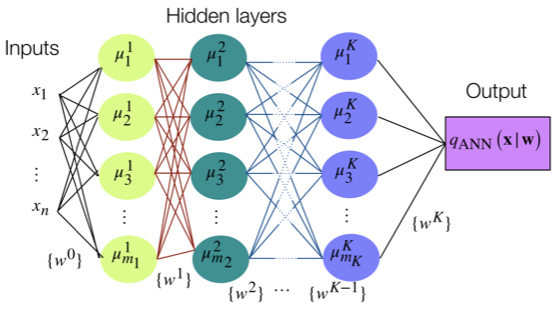}\caption{Schematic representation of the ANN used for deep learning. \label{fig:ANN}}
\end{figure}

We write the unknown reaction coordinate  $q\left(\mathbf{x}\right) $  in terms of the following artificial neural network (ANN):
\begin{equation}
\begin{split}q\left(\mathbf{x}\right) & =q_{\mathrm{ANN}}\left(\mathbf{x}|\mathbf{w}\right)\\
\equiv & \sum_{j_{K}=1}^{m_{K}}h\left(\ldots h\left\{ \sum_{j_{2}=1}^{m_{2}}h\left[\sum_{j_{1}=1}^{m_{1}}h\left(\sum_{\alpha=1}^{n}x_{\alpha}w_{\alpha,j_{1}}^{0}+w_{0,j_{1}}^{0}\right)w_{j_{1},j_{2}}^{1}+w_{0,j_{2}}^{1}\right]w_{j_{2},j_{3}}^{2}+w_{0,j_{3}}^{2}\right\} \ldots\right)w_{j_{K}}^{K}+w_{0}^{K},
\end{split}
\label{eq:ANN-1}
\end{equation}
which represents a network (Figure \ref{fig:ANN}) that takes $x_{1,}x_{2},\ldots,x_{n}$
different inputs, with $K$ hidden layers containing each $m_{1},m_{2},\ldots,m_{K}$
different nodes. The weight matrix $\mathbf{w}$, i.e., the fitting
parameters, defines the connections between nodes, with $w_{ik}^{j}$
connecting node $i$ of layer $j$ with node $k$ of layer $j+1$.
The ``activation'' function $h$ is a non linear function mimicking
the threshold firing behavior of biological neurons. 

\section{Symbolic Regression}

After determining the most relevant inputs we use symbolic regression, and in particular differentiable Cartesian genetic programming, to approximate the trained ANN with a simple expression \cite{Izzo2016}. We employ a 1+4 evolutionary strategy  for 250 generations
where every change in the genome of an offspring is followed by 2500
Newton steps in the weight space of that expression. We add a regularization
term $\lambda n $ where $n$ is proportional to the number of active genes to avoid overfitting. We test regularization
values of $\lambda \in \left[ 0.005, 0.001,  0.01 \right]$.

\begin{figure}
  \includegraphics[width=0.70\columnwidth]{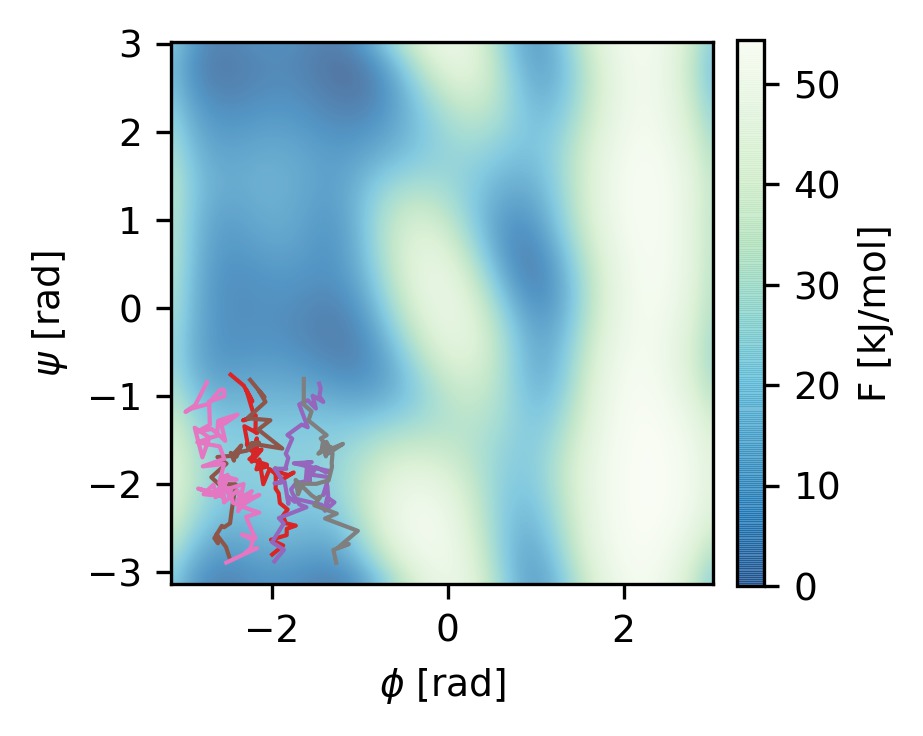}\caption{Free energy surface of alanine dipeptide as a function of the main dihedral angles $\phi$ and $\psi$. In the lower left corner we report some representative TPs that cross the high energy barrier, which correspond  to a rotation of $\psi$. The free energy was obtained by using metadynamics. \label{fig:ala2FES}}
\end{figure}

\begin{figure}
  \includegraphics[width=1.0\columnwidth]{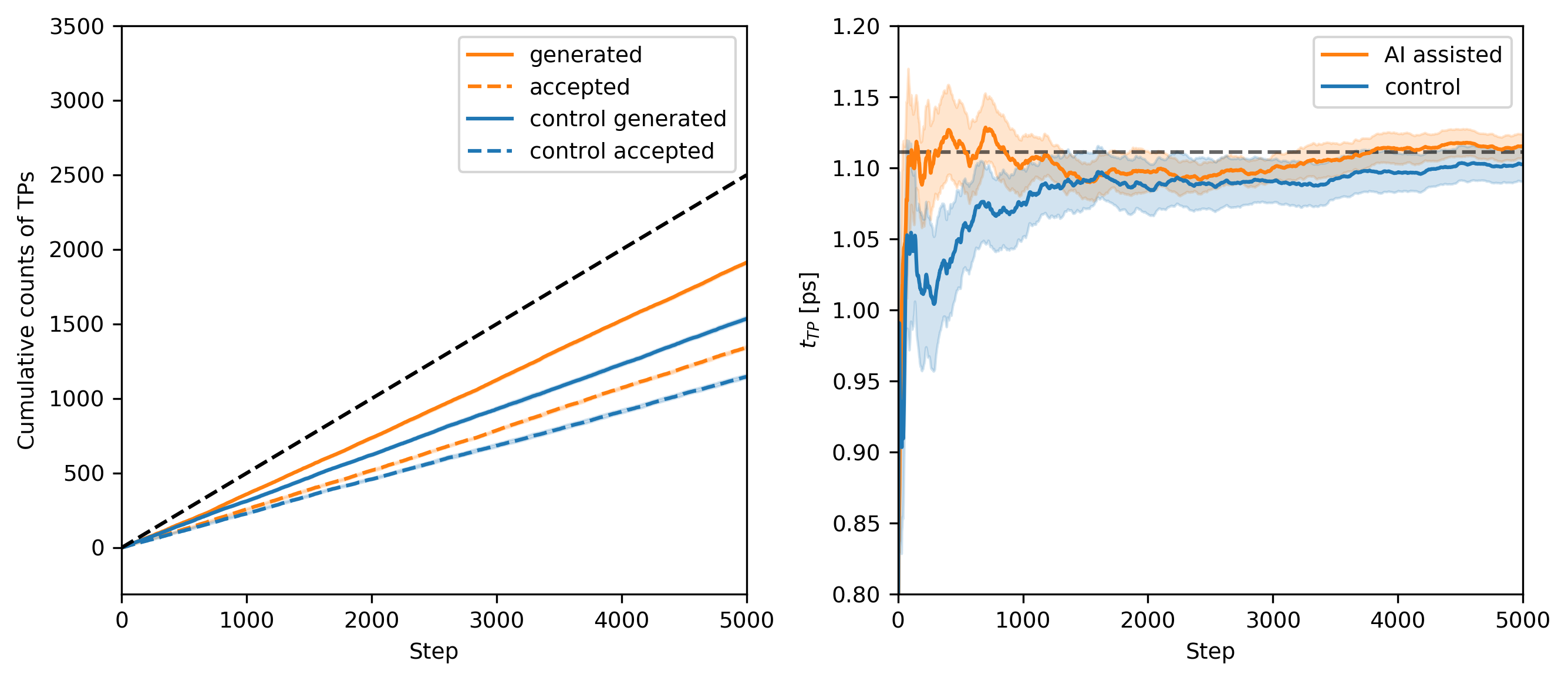}\caption{TP sampling results for alanine dipeptide. (Left) Cumulative count of
  generated (continuous line) and accepted (dashed line) TPs in
  AI-assisted MD simulations (orange) and standard control Trantision Path Sampling (TPS) (blue).
   Each line shows the average over 10
  independent runs. The standard error of the mean is smaller than the
  thickness of the lines. The black dashed line represent the
  theoretical maximum for generation of TPs. (Right) Convergence of TP time. Running average for AI-assisted MD simulations (orange) and control TPS simulation (blue). Each line shows the average over 10
  independent runs. The shaded area is given by the standard error of the mean. The black dashed line represent the
  reference value calculated in a long control TPS simulation. \label{fig:ala2FES}}
\end{figure}

\begin{table}
	\centering
	\caption[Alanine dipeptide: Relevant coordinates for reduced symmetry functions]{Input relevance analysis for alnine dipeptide.  The first 4 most relevant input coordinates and their range of values. All symmetry functions probe the oxygens of water using the parameters in brackets, the subscript to $G$ indicates the index of the central atom. Index 5 is the oxygen of the acetyl involved in the peptide bond. Index 12 is the 3 symmetry related hydrogen atom of the methyl-group of alanine and index 16 is the nitrogen of the methylamide.}
	\label{tab:ala_important_coords_sf_red}
	\begin{tabular}{lcc}
			\toprule Index & Definition and range & Normalized relevance $r\left( x_i \right)$ \\
			$x_1$ & $\tilde{\psi} = 0.5(\cos(\psi + \pi) + 1) \in [0.1306, 0.9778]$ & $1.00$\\
			$x_2$ & $G^5_{5}(\eta=40, r_s=0.175, \zeta=64, \lambda=-1) \in [0.0032, 1.7550]$ & $0.07$\\
			$x_3$ & $G^5_{12}(\eta=40, r_s=0.175, \zeta=64, \lambda=-1) \in [0.0035, 1.3719]$ & $0.05$\\
			$x_4$ & $G^5_{16}(\eta=40, r_s=0.175, \zeta=64, \lambda=-1) \in [0.0053, 1.5534]$ & $0.05$\\
			\bottomrule
	\end{tabular}
\end{table}

\begin{table}
	\centering
	\caption[Alanine dipeptide: Symbolic regression results for reduced symmetry functions]{Symbolic regression results for alanine dipeptide. 
	We used as input the first or the four most important coordinates as shown in Table I. Due to the regularization we applied, some input coordinates are not part of the final converged expressions. We used every 10th shooting configurations extracted from the TPS simulation as training set for the symbolic regression.  The loss values $L_{\mathrm{ANN}}$ and $L_{\mathrm{SR}}$ were calculated for the reduced set of points and correspond to the test loss per shooting configuration of the ANN  and symbolic regression expression respectively. For every parameter combination the symbolic regression was repeated three times to quantify the stability of the results.}
	\label{tab:ala_SR_sf_red}
		\begin{tabular}{cccccl}
		\toprule
		$L_{\mathrm{ANN}}$ & \makecell{Selected\\coordinates} & $\lambda$ & $L_{\mathrm{SR}}$ & Frequency & Final expression \\
		\midrule
		\multirow{9}{*}{1.031} & \multirow{2}{*}{1} & 0.01 & 1.057 &  3/3 & {$\!\begin{aligned}
			q_{\mathrm{SR}} &= -5.241 \tilde{\psi} - 1.865\ln(\tilde{\psi}) + 2.550
			\end{aligned}$}\\
		\cmidrule{3-6}
		& & 0.005 & 1.057 & 3/3 & {$\!\begin{aligned}
			q_{\mathrm{SR}} &= -5.241 \tilde{\psi} - 1.865\ln(\tilde{\psi}) + 2.550
			\end{aligned}$}\\
		\cmidrule{3-6}
		& & \multirow{2}{*}{0.001} & 1.057 & 2/3 & {$\!\begin{aligned}
			q_{\mathrm{SR}} &= -5.241 \tilde{\psi} - 1.865\ln(\tilde{\psi}) + 2.550
			\end{aligned}$}\\
		& & & 1.049 & 1/3 & {$\!\begin{aligned}
			q_{\mathrm{SR}} &= -8.377\tilde{\psi} + 5.424
			\end{aligned}$}\\
		\cmidrule{2-6}
		& \multirow{5}{*}{4} & \multirow{2}{*}{0.01} & 1.051 & 2/3 & {$\!\begin{aligned}
			q_{\mathrm{SR}} &= -8.384 \tilde{\psi} + 5.622\exp(-0.285x_3)
			\end{aligned}$}\\
		& & & 1.053 &  1/3 & {$\!\begin{aligned}
			q_{\mathrm{SR}} &= -1.438 x_3 - 5.153\ln(\tilde{\psi}) - 2.172
			\end{aligned}$}\\
		\cmidrule{3-6}
		& & 0.005 & 1.051 & 3/3 & {$\!\begin{aligned}
			q_{\mathrm{SR}} &= -8.384 \tilde{\psi} + 5.622\exp(-0.285x_3)
			\end{aligned}$}\\
		\cmidrule{3-6}
		& & \multirow{2}{*}{0.001} & 1.051 & 2/3 & {$\!\begin{aligned}
			q_{\mathrm{SR}} &= -8.384 \tilde{\psi} + 5.622\exp(-0.285x_3)
			\end{aligned}$}\\
		& & & 1.053 & 1/3 & {$\!\begin{aligned}
			q_{\mathrm{SR}} &= -1.438 x_3 - 5.153\ln(\tilde{\psi}) - 2.172
			\end{aligned}$}\\
		\bottomrule
	\end{tabular}
\end{table}

\begin{figure}
  \includegraphics[width=1.0\columnwidth]{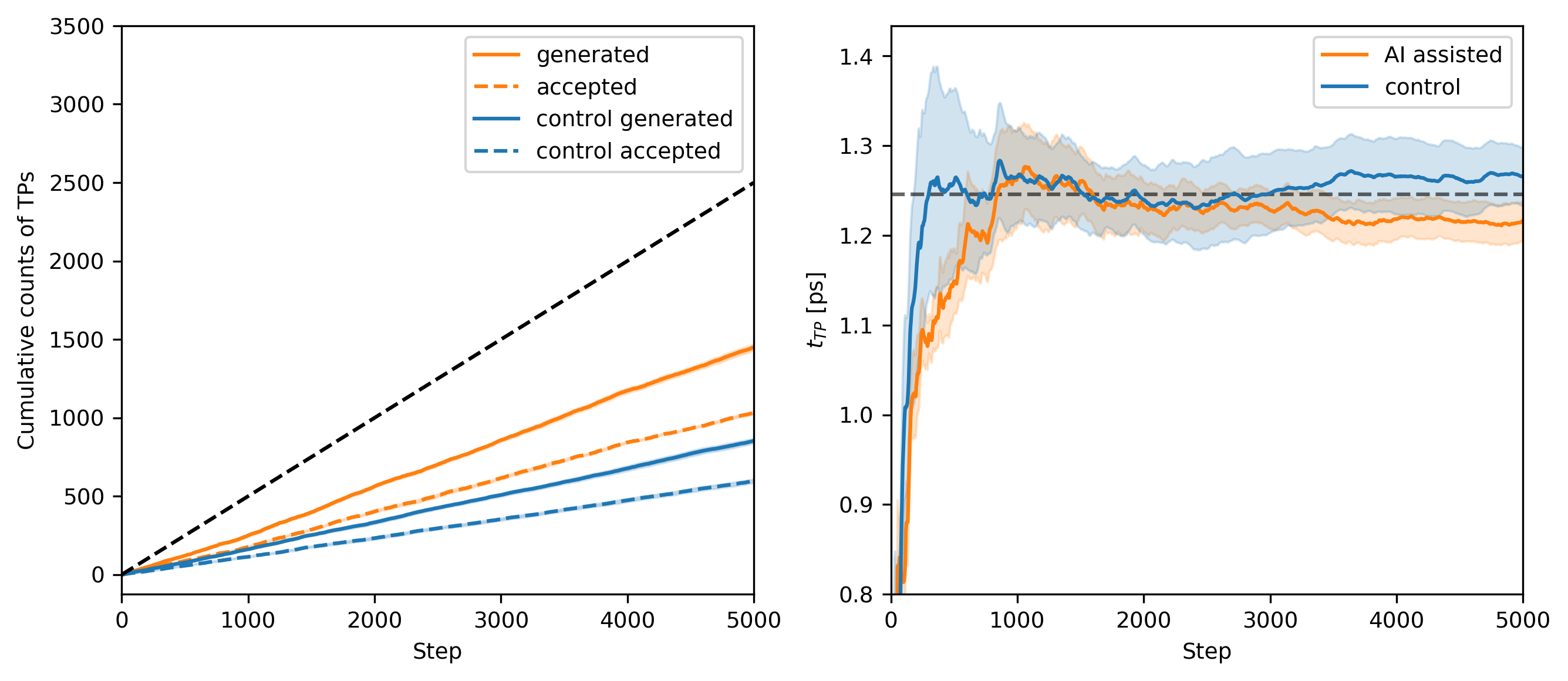}\caption{TP sampling results for LiCl. (Left) Cumulative count of
  generated (continuous line) and accepted (dashed line) TPs in
  AI-assisted MD simulations (orange) and standard control TPS (blue).
   Each line shows the average over 10
  independent runs. The standard error of the mean is smaller than the
  thickness of the lines. The black dashed line represent the
  theoretical maximum for generation of TPs. (Right) Convergence of TP time. Running average for AI-assisted MD simulations (orange) and control TPS simulation (blue). Each line shows the average over 10
  independent runs. The shaded area is given by the standard error of the mean. The black dashed line represent the
  reference value calculated in a long control TPS simulation. \label{fig:ala2FES}}
\end{figure}

\begin{table}
	\centering
	\caption[LiCl: Relevant coordinates]{Input relevance analysis for LiCl. Range of values of the most relevant coordinates. The symmetry functions are centered on the central atom indicated by the subscript and use the parameters in brackets to probe for the solvent species in square brackets. }
	\label{tab:licl_important_coords_sf}
		\begin{tabular}{lcc}
			\toprule Index & Definition and range & Normalized relevance $r\left( x_i \right)$ \\
			$x_1$  & $G^5_{\mathrm{Li}}(\eta=160, r_s=0.7, \zeta=4, \lambda=-1)[\mathrm{Cl}] \in [0.0, 0.491]$ & $1.00$\\
			$x_2$  & $G^5_{\mathrm{Li}}(\eta=160, r_s=0.7, \zeta=64, \lambda=-1)[\mathrm{Cl}] \in [0.0, 0.417]$ & $0.78$ \\
			$x_3$  & $G^5_{\mathrm{Li}}(\eta=160, r_s=0.625, \zeta=16, \lambda=-1)[\mathrm{Cl}] \in [0.0, 0.007]$ & $0.68$\\
			$x_4$  & $G^5_{\mathrm{Li}}(\eta=160, r_s=0.625, \zeta=4, \lambda=-1)[\mathrm{Cl}] \in [0.0, 0.343]$ & $0.45$\\
			$x_5$  & $G^5_{\mathrm{Cl}}(\eta=160, r_s=0.7, \zeta=4, \lambda=1)[\mathrm{Li}] \in [0.0, 6.940]$ & $0.44$ \\
			$x_6$  & $G^2_{\mathrm{Cl}}(\eta=300, r_s=0.4)[\mathrm{Li}] \in [0.0, 5.072]$ & $0.38$\\
			$x_7$  & $G^5_{\mathrm{Cl}}(\eta=160, r_s=0.7, \zeta=2, \lambda=1)[\mathrm{Li}] \in [0.0, 5.678]$ & $0.32$\\
			$x_8$  & $G^5_{\mathrm{Cl}}(\eta=160, r_s=0.625, \zeta=4, \lambda=-1)[\mathrm{Li}] \in [0.0,~ 0.971]$ & $0.27$\\
			$x_9$  & $G^5_{\mathrm{Li}}(\eta=160, r_s=0.175, \zeta=16, \lambda=-1)[\mathrm{O}] \in [0.061,~ 6.713]$ & $0.23$\\
			$x_{10}$ & $G^5_{\mathrm{Cl}}(\eta=160, r_s=0.475, \zeta=4, \lambda=-1)[\mathrm{Li}] \in [0.0, 7.666]$ & $0.19$\\
			$x_{69}$ & $r_{\mathrm{LiCl}} \in [0.240, 0.470]$ & $0.006$\\
			\bottomrule
	\end{tabular}
\end{table}

\begin{table}[bt]
	\centering
	\caption[LiCl: Symbolic regression results for symmetry functions and $r_{\mathrm{LiCl}}$]{Symbolic regression results for LiCl.
	We use as input the 3 or 10 most important coordinates and the interionic distance as shown in Table III. Due to the regularization we apply, some input coordinates 	are not part of the final converged expressions. We used every 10th shooting configurations extracted from the TPS simulation as training set for the symbolic regression.  The loss values $L_{\mathrm{ANN}}$ and $L_{\mathrm{SR}}$ are calculated for the reduced set of points and correspond to the test loss per shooting point of the ANN  and symbolic regression expression respectively. For every parameter combination the symbolic regression was repeated three times to quantify the stability of the results.}
	\label{tab:licl_SR_sf_dist}
	\begin{tabular}{cccccl}
		\toprule
		$L_{\mathrm{ANN}}$ & \makecell{Selected\\coordinates} & $\lambda$ & $L_{\mathrm{SR}}$ & Frequency & Final expression \\
		\midrule
		\multirow{13}{*}{0.706} & \multirow{5}{*}{3 + $r_{\mathrm{LiCl}}$} & \multirow{1}{*}{0.01} & 0.757 &  3/3 & {$\!\begin{alignedat}{1}
			q_{\mathrm{SR}} &= 29.837 r_{\mathrm{LiCl}} - 8.485\exp(-0.728 x_1)
			\end{alignedat}$}\\
		\cmidrule{3-6}
		& & \multirow{1}{*}{0.005} & 0.757 &  3/3 & {$\!\begin{alignedat}{1}
			q_{\mathrm{SR}} &= 29.837 r_{\mathrm{LiCl}} - 8.485\exp(-0.728 x_1)
			\end{alignedat}$}\\
		\cmidrule{3-6}
		& & \multirow{3}{*}{0.001} & 0.757 &  1/3 & {$\!\begin{alignedat}{1}
			q_{\mathrm{SR}} &= 29.837 r_{\mathrm{LiCl}} - 8.485\exp(-0.728 x_1)
			\end{alignedat}$}\\
		& & & 0.754 & 1/3 & {$\!\begin{alignedat}{1}
			q_{\mathrm{SR}} &= 29.896 r_{\mathrm{LiCl}} - 8.518\exp(-0.617 x_1)\\
			&+ 4.953 x_2
			\end{alignedat}$}\\
		& & & 0.746 & 1/3 & {$\!\begin{alignedat}{1}
			q_{\mathrm{SR}} &= 30.115 r_{\mathrm{LiCl}}\exp(0.996 x_2)\\
			&- 8.973\sin(1.260\exp(3.269 x_1))
			\end{alignedat}$}\\
		\cmidrule{2-6}
		& \multirow{8}{*}{10 + $r_{\mathrm{LiCl}}$} & \multirow{3}{*}{0.01} & 0.753 & 1/3 & {$\!\begin{alignedat}{1}
			q_{\mathrm{SR}} &= 29.592 r_{\mathrm{LiCl}} - 8.204\exp(0.028 x_9)\\
			&+ 10.093 x_4
			\end{alignedat}$}\\
		& & &  0.757 & 1/3 & {$\!\begin{alignedat}{1}
			q_{\mathrm{SR}} &= 29.837 r_{\mathrm{LiCl}} - 8.405\exp(-0.728 x_1)
			\end{alignedat}$}\\
		& & &  0.761 & 1/3 & {$\!\begin{alignedat}{1}
			q_{\mathrm{SR}} &= 28.993 r_{\mathrm{LiCl}} - 0.182\ln(x_9) - 8.273
			\end{alignedat}$}\\
		\cmidrule{3-6}
		& & \multirow{2}{*}{0.005} & 0.758 & 2/3 & {$\!\begin{alignedat}{1}
			q_{\mathrm{SR}} &= 29.913 r_{\mathrm{LiCl}} - 8.473\exp(-1.158 x_4)
			\end{alignedat}$}\\
		& & & 0.757 & 1/3 & {$\!\begin{alignedat}{1}
			q_{\mathrm{SR}} &= 29.837 r_{\mathrm{LiCl}} - 8.485\exp(-0.728 x_1)
			\end{alignedat}$}\\
		\cmidrule{3-6}
		& & \multirow{3}{*}{0.001} & 0.751 & 1/3 & {$\!\begin{alignedat}{1}
			q_{\mathrm{SR}} &= 29.510 r_{\mathrm{LiCl}} - 8.187\exp(-0.866 x_1)\\
			&- 0.289 x_9
			\end{alignedat}$}\\
		& & & 0.751 & 1/3 & {$\!\begin{alignedat}{1}
			q_{\mathrm{SR}} &= 29.550 r_{\mathrm{LiCl}} - 8.214\exp(0.031 x_9)\\
			&+ 6.767 x_1
			\end{alignedat}$}\\
		& & & 0.757 & 1/3 & {$\!\begin{alignedat}{1}
			q_{\mathrm{SR}} &= 29.837 r_{\mathrm{LiCl}} - 8.485\exp(-0.728 x_1)
			\end{alignedat}$}\\
		\bottomrule
	\end{tabular}
\end{table}

\bibliographystyle{apsrev4-1}
\bibliography{biblio}